\documentclass{jfm}
\usepackage{color}

\begin{document}

\newtheorem{lemma}{Lemma}
\newtheorem{corollary}{Corollary}

\shorttitle{Rotations of inertialess triaxial ellipsoids in turbulence} 
\shortauthor{N. Pujara \& E. Variano} 

\title{Rotations of small, inertialess triaxial ellipsoids in isotropic turbulence}

\author
 {
 Nimish Pujara\aff{1}
  \corresp{\email{pujara@berkeley.edu}},
  Evan Variano\aff{1}
  }

\affiliation
{
\aff{1}
Department of Civil and Environmental Engineering, University of California, Berkeley, CA 94720, USA
}

\maketitle

\begin{abstract} 
The statistics of rotational motion of small, inertialess triaxial ellipsoids are computed along Lagrangian trajectories extracted from direct numerical simulations of homogeneous isotropic turbulence. The particle angular velocity and its components along the three principal axes of the particle are considered, expanding on the results presented by \citet{ChevillardMeneveau13}. The variance of the particle angular velocity, referred to as the particle enstrophy, is found to increase for particles with elongated shapes. This trend is explained by considering the contributions of vorticity and strain-rate to particle rotation. It is found that the majority of particle enstrophy is due to fluid vorticity. Strain-rate-induced rotations, which are sensitive to shape, are mostly cancelled by strain-vorticity interactions. The remainder of the strain-rate-induced rotations are responsible for weak variations in particle enstrophy. For particles of all shapes, the majority of the enstrophy is in rotations about the longest axis, which is due to alignment between the longest axis and fluid vorticity. The integral timescale for particle angular velocities about each axis reveals that rotations are most persistent about the longest axis, but that a full revolution is rare.
\end{abstract}


\section{Introduction}

\citet{Jeffery22} derived the leading-order equations of motion in the limit of low Reynolds number for a triaxial ellipsoid that is of the same density as the surrounding fluid. The particle translates with the same velocity as the fluid that it displaces, but its rotation differs from the displaced fluid and is governed the local velocity gradients. The applicability of these equations has been verified experimentally \citep{Taylor23, TrevelyanMason51, AnczurowskiMason68, Parsa11, Parsa12, Einarsson13, Ni15} and many studies have employed them to study the behaviour of anisotropic particles in laminar and turbulent flows \citep{ShinKoch05, Parsa11, ChevillardMeneveau13, Gustavsson14, Byron15, Challabotla15, Zhao15} with applications such as paper manufacturing \citep{Lundell11}, microbial and plankton ecology \citep{PedleyKessler92, Guasto12}, and sediment transport.

While most studies restrict themselves to axisymmetric ellipsoids, \textit{i.e.}, oblate and prolate spheroids, \citet{ChevillardMeneveau13} examined triaxial ellipsoids. They presented statistics of the rate of change of orientation vectors and used these to evaluate stochastic models of the Lagriangian velocity gradient tensor. We are motivated by the desire to understand the interaction of planktonic organisms with their turbulent flow environment. Since a variety of life functions depend on organismal rotation \citep{Kiorboe2008, Guasto12}, it is important to understand how shape affects particle angular velocity, its partioning amongst the particle's axes, the likelihood of extreme events, and its correlation timescales. In this paper, we address these gaps in the literature. We start in section 2 by introducing the particle-shape parameter space, reviewing the equations for particle rotational motion, and providing methods to solve these equations in turbulent flow. Section 3 presents the statistics of rotational motion of triaxial ellipsoids in homogenous isotropic turbulence and discusses the trends of these statistics with variation in particle shape. Section 4 provides the conclusions.


\section{Methods}\label{sec:methods}

\subsection{Triaxial ellipsoids}\label{sec:triaxials}

\begin{figure}
  \centerline{\includegraphics[width=12 cm]{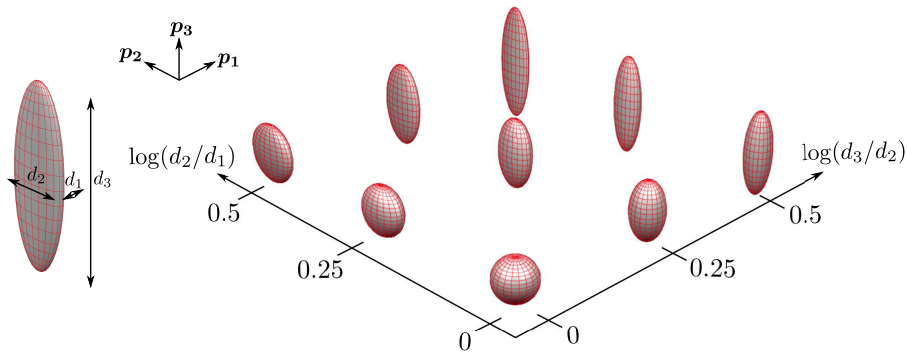}}
  \caption{\textit{Left}: Nomenclature for triaxial ellipsoids: $d_{i}$ are diameters corresponding to particle orientation vectors $\boldsymbol{p_{i}}$ for $i=1,2,3$ labelled such that $d_{3} > d_{2} > d_{1}$. \textit{Right}: Triaxial ellipsoids showing part of the parameter space to be explored.}
\label{fig:triaxial_ellipsoid}
\end{figure}

\noindent A diagram of triaxial ellipsoid shape parameters is shown in figure \ref{fig:triaxial_ellipsoid}. We label the three particle axes such that $d_{3}>d_{2}>d_{1}$ and use the ratios $d_{3}/d_{2}$ and $d_{2}/d_{1}$ to describe particle shape. The degree to which a particle is elongated is indicated by $d_{3}/d_{2}$ and the degree to which a particle is flattened is indicated by $d_{2}/d_{1}$. Other definitions of the parameter space are also possible \citep[\textit{e.g.},][]{Dusenbery09, ChevillardMeneveau13}, but in our definition, a particle of a given shape holds a unique position in the particle-shape parameter space.

We explore the particle shapes in the space spanned by $\left[\log(d_{3}/d_{2}),\log(d_{2}/d_{1})\right] \in \left(0 \rightarrow 1,0 \rightarrow 1\right)$ while keeping the particle volume constant ($d_{3}d_{2}d_{1}=\textrm{const.}$). Particles whose shapes are beyond the boundaries of our parameter space are not expected to show qualitatively different dynamics than those at the boundaries, as suggested by previous results. Specifically, work on the statistics of spheroid rotations have shown that the majority of variation in the statistics of spheroid rotation in turbulent flows due to a change of shape occurs over a range of aspect ratio $10^{-1} \rightarrow 10^{1}$ \citep{Parsa12, Byron15}. 

\begin{figure}
  \centerline{\includegraphics[width=10 cm]{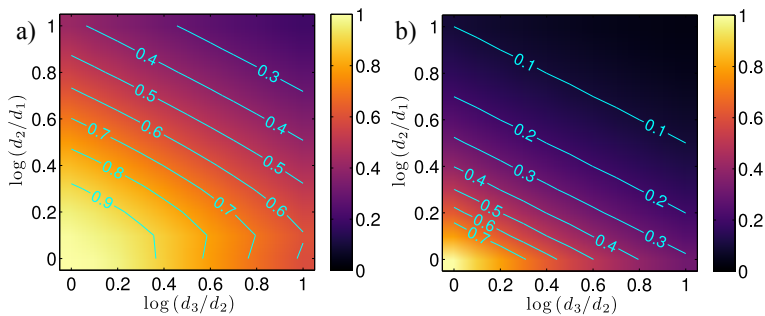}}
  \caption{Single-parameter shape descriptions shown within the two-parameter shape space defined in figure \ref{fig:triaxial_ellipsoid}. a) Wadell sphericity parameter, $\Psi$; b) Corey shape factor, $\Phi$.}
\label{fig:sphericity}
\end{figure}

In applications such as sediment transport, shape effects are commonly described by two single-parameter measures of particle sphericity; these are the Wadell sphericity paramter, $\Psi$ \citep{Wadell32}, and the Corey shape factor, $\Phi$ \citep[\textit{e.g.},][]{Dietrich82}. Figure \ref{fig:sphericity} shows how $\Psi$ and $\Phi$ are related to the particle-shape parameter space considered here. $\Psi$ is defined as the ratio of the surface area of a volume-matched sphere to the surface area of a particle and $\Phi$ is defined as the ratio of the maximum cross-sectional area of a volume-matched sphere to the maximum cross-sectional area of a particle. Using the expression for the surface area of a triaxial ellipsoid, the inverse of Wadell's sphericity parameter is given by 

\begin{equation}
{\Psi}^{-1} = \frac{1}{2} \left(\frac{d_{3}/d_{2}}{d_{2}/d_{1}}\right)^{1/3} \left\{ {\left(\frac{d_{3}}{d_{1}}\right)}^{-1} + {\left(\frac{d_{2}}{d_{1}}\right)} \frac{\left[E(\phi,k) \sin^{2}\phi + F(\phi,k) \cos^{2}\phi \right]}{\sin \phi} \right\},
\label{eq:asphericity_ellipsoids} 
\end{equation}

\noindent where

\begin{equation}
\cos \phi = \frac{d_{1}}{d_{3}} \; ; \; k^{2} = {\left(\frac{d_{3}}{d_{2}}\right)}^2 \frac{(d_{2}/d_{1})^2 - 1}{(d_{3}/d_{1})^2 - 1},
\end{equation}

\noindent and $F(\phi,k)$ and $E(\phi,k)$ are the incomplete elliptical integrals of the first kind and second kind, respectively. The Corey shape factor for triaxial ellipsoids is given by

\begin{equation}
\Phi =\left(\frac{d_{3}}{d_{2}}\right)^{-1/2}\left(\frac{d_{2}}{d_{1}}\right)^{-1}.
\end{equation}

\subsection{Evolution of particle orientation}

\noindent We track the evolution of a particle's orientation vectors, $(\boldsymbol{p_{1}}, \boldsymbol{p_{2}}, \boldsymbol{p_{3}})$, in a fixed (unmoving) laboratory reference frame. The time rate of change of the orientation vectors, ($\boldsymbol{\dot{p_{1}}}, \boldsymbol{\dot{p_{2}}}, \boldsymbol{\dot{p_{3}}}$), is related to the angular velocity of the particle, $\boldsymbol{\omega_{p}}$, by the following kinematic relations:

\begin{subeqnarray}
\boldsymbol{\dot{p_{1}}} = \boldsymbol{\omega_{p}} \times \boldsymbol{{p_{1}}}, \\
\boldsymbol{\dot{p_{2}}} = \boldsymbol{\omega_{p}} \times \boldsymbol{{p_{2}}}, \\
\boldsymbol{\dot{p_{3}}} = \boldsymbol{\omega_{p}} \times \boldsymbol{{p_{3}}}.
\label{eq:kinematic_rotation}
\end{subeqnarray}

\noindent In the limit of small, inertialess triaxial ellipsoids immersed in a viscous fluid, \citet{Jeffery22} provided the following equations to describe the rates of rotation of a particle about each of its axes in terms of the local gradients of velocity:

\begin{subeqnarray}
  \boldsymbol{\omega_{p}}\bcdot\boldsymbol{p_{1}} = \frac{1}{2}\boldsymbol{\omega}\bcdot\boldsymbol{p_{1}} + \lambda_{1}\left( {\boldsymbol{p_{2}}}^{T}\boldsymbol{S}\boldsymbol{p_{3}}\right), \\
\boldsymbol{\omega_{p}}\bcdot\boldsymbol{p_{2}} = \frac{1}{2}\boldsymbol{\omega}\bcdot\boldsymbol{p_{2}} + \lambda_{2}\left( {\boldsymbol{p_{3}}}^{T}\boldsymbol{S}\boldsymbol{p_{1}}\right),\\
\boldsymbol{\omega_{p}}\bcdot\boldsymbol{p_{3}} = \frac{1}{2}\boldsymbol{\omega}\bcdot\boldsymbol{p_{3}} + \lambda_{3}\left( {\boldsymbol{p_{1}}}^{T}\boldsymbol{S}\boldsymbol{p_{2}}\right).
\label{eq:jeffery}
\end{subeqnarray}

\noindent In the above equations, $\boldsymbol{\omega} = \bnabla \times \boldsymbol{u}$ is the fluid vorticity, and $\boldsymbol{S} = (1/2)[\nabla \boldsymbol{u} + (\nabla \boldsymbol{u})^T]$ is the strain-rate tensor given by the symmetric part of the velocity gradient tensor. The symmetric and anti-symmetric parts of the velocity gradient tensor are denoted as $\nabla \boldsymbol{u} = {\boldsymbol{S}} + {\boldsymbol{\Omega}}$. The shape parameters $\lambda_{i}$ are given by:

\begin{equation}
\lambda_{1} = \frac{(d_{2}/d_{3})^2 - 1}{(d_{2}/d_{3})^2 + 1}; \quad \lambda_{2} = \frac{(d_{3}/d_{1})^2 - 1}{(d_{3}/d_{1})^2 + 1}; \quad \lambda_{3} = \frac{(d_{1}/d_{2})^2 - 1}{(d_{1}/d_{2})^2 + 1}.
  \label{eq:lambda}
\end{equation}

\noindent \citeauthor{Jeffery22}'s (\citeyear{Jeffery22}) equations, as presented in Eq. (\ref{eq:jeffery}), are in a fixed (unmoving) laboratory reference frame. Combining the rotation of the particle about all three of its axes, the total angular velocity of the particle in the fixed reference frame is given by

\begin{equation}
\boldsymbol{\omega_{p}} = \frac{1}{2}\boldsymbol{\omega} + \lambda_{1} \boldsymbol{p_{1}} \left( {\boldsymbol{p_{2}}}^{T}\boldsymbol{S}\boldsymbol{p_{3}}\right) + \lambda_{2} \boldsymbol{p_{2}} \left( {\boldsymbol{p_{3}}}^{T}\boldsymbol{S}\boldsymbol{p_{1}}\right) + \lambda_{3} \boldsymbol{p_{3}} \left({\boldsymbol{p_{1}}}^{T}\boldsymbol{S}\boldsymbol{p_{2}}\right).
\label{eq:omega_p}
\end{equation}

\noindent Inserting Eq. (\ref{eq:omega_p}) into Eq. (\ref{eq:kinematic_rotation}), and employing the identity $\frac{1}{2} \left(\boldsymbol{\omega} \times \boldsymbol{p}\right) \equiv \boldsymbol{\Omega \boldsymbol{p}}$, the rate of change of orientation vectors can be written as

\begin{subeqnarray}
  \dot{\boldsymbol{p_{1}}} = \boldsymbol{\Omega \boldsymbol{p_{1}}} - \lambda_{2} \boldsymbol{p_{3}} \left( {\boldsymbol{p_{3}}}^{T}\boldsymbol{S}\boldsymbol{p_{1}}\right) + \lambda_{3} \boldsymbol{p_{2}} \left({\boldsymbol{p_{1}}}^{T}\boldsymbol{S}\boldsymbol{p_{2}}\right),\\
  \dot{\boldsymbol{p_{2}}} = \boldsymbol{\Omega \boldsymbol{p_{2}}} + \lambda_{1} \boldsymbol{p_{3}} \left( {\boldsymbol{p_{2}}}^{T}\boldsymbol{S}\boldsymbol{p_{3}}\right) - \lambda_{3} \boldsymbol{p_{1}} \left({\boldsymbol{p_{1}}}^{T}\boldsymbol{S}\boldsymbol{p_{2}}\right),\\
\dot{\boldsymbol{p_{3}}} = \boldsymbol{\Omega \boldsymbol{p_{3}}} - \lambda_{1} \boldsymbol{p_{2}} \left( {\boldsymbol{p_{2}}}^{T}\boldsymbol{S}\boldsymbol{p_{3}}\right) + \lambda_{2} \boldsymbol{p_{1}} \left( {\boldsymbol{p_{3}}}^{T}\boldsymbol{S}\boldsymbol{p_{1}}\right).
  \label{eq:pdot_evolution}
\end{subeqnarray}

\noindent The above equations are the same as those derived by \citet{JunkIllner07}, and used by \citet{ChevillardMeneveau13} and \citet{Einarsson15}. The following section describes how these equations were numerically integrated along particle trajectories in turbulent flow.

\subsection{Direct numerical simulations}

\noindent Data from direct numerical simulations (DNS) of homogeneous isotropic turbulence available from the Johns Hopkins University turbulence database \citep{Perlman07, Li08} are used to calculate the evolution of particle orientation along particle trajectories. The dataset consists of forced isotropic turbulence on a triply periodic cube of $1024^{3}$ points over 1024 time samples that span roughly one large-eddy turnover time. The Taylor micro-scale Reynolds number is ${\Rey_{\lambda}} \approx 433$ and the ratio of large-eddy turnover time (integral timescale), $T$, to the Kolmogorov timescale, $\tau_{\eta}$, is given by $T \approx 45\tau_{\eta}$. 

Timeseries of the velocity gradient tensor along 7500 Lagrangian trajectories are extracted and used to integrate the equations for particle orientation, Eq. (\ref{eq:pdot_evolution}). The duration of each trajectory is approximately one large-eddy turnover time. The Lagrangian tracking is done using a second-order Runge-Kutta scheme with a time-step given by $\Delta t_{p}/\tau_{\eta} = 0.009$, which is a factor of five smaller than the time-step stored on the dataset. Spatial interpolation is done using sixth-order Lagrange polynomials and interpolation in time is done using cubic Hermite interpolation. Further details of the Lagrangian tracking algorithm can be found in \citet{Yu12}. Using the Lagrangian velocity gradient tensor timeseries, Eqs. (\ref{eq:pdot_evolution}) are numerically integrated using a fourth-order Runge-Kutta scheme. Particles are initialized with random orientations and random positions.


\section{Results and discussion}\label{sec:results}

\subsection{Initial transient period for particle orientation}

\begin{figure}
  \centerline{\includegraphics[width=7 cm]{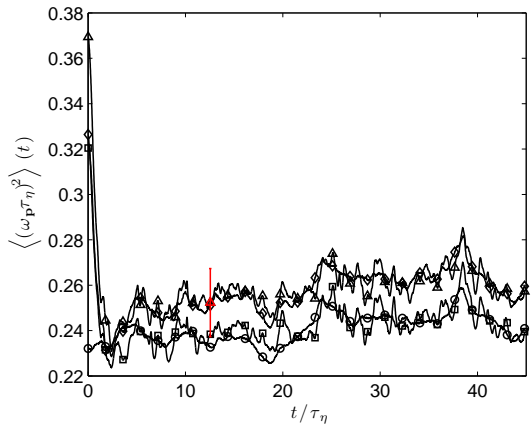}}
  \caption{Variance of particle angular velocity computed at each time-step from the ensemble of 7500 particles for four particle shapes: \protect$\circ$, sphere ($d_{3}/d_{2}=d_{2}/d_{1}=1$); \protect$\square$, disc ($d_{3}/d_{2}=1,d_{2}/d_{1}=10$); \protect$\triangle$, rod ($d_{3}/d_{2}=10,d_{2}/d_{1}=1$); \protect$\diamond$, triaxial ($d_{3}/d_{2}=d_{2}/d_{1}=10$). The example error bar in red at $t/\tau_{\eta} \approx 12.5$ shows the 95\% confidence intervals calculated using data only at that time-step for the triaxial case.}
\label{fig:transient}
\end{figure}

\noindent At $t=0$, particles are oriented randomly and there is no correlation between the particle orientation vectors, $(\boldsymbol{p_{1}},\boldsymbol{p_{2}},\boldsymbol{p_{3}})$, and the local velocity gradient tensor, $ \nabla \boldsymbol{u} = {\boldsymbol{S}} + {\boldsymbol{\Omega}}$. Thus, the statistics of particle rotation at $t=0$ can be predicted in terms of the Kolmogorov timescale for isotropic turbulence \citep{ShinKoch05,Parsa13,ChevillardMeneveau13}. With increasing time, particle orientations start to become aligned with the local velocity gradient tensor and the statistics of particle rotation evolve. Figure \ref{fig:transient} shows the time evolution of the dimensionless variance of particle angular velocity for four different shapes that occupy the four corners of the particle-shape parameter space. The initial variance of particle angular velocity is higher than its steady-state value, which is reached by $t \approx 5 \tau_{\eta}$. All subsequent results in this paper are calculated from an ensemble average over all particles and over all timesteps for $6\tau_{\eta}\le t\le 45\tau_{\eta}$. Neglecting data from the intial transient period removes the influence of the intial conditions on the statistics of particle rotation despite the fact that the equations of motion are derived in the limit of small Reynolds number and small Stokes number \citep{Jeffery22}.

\subsection{Variance and kurtosis of particle angular velocities}

\noindent The angular velocity of a particle is the vector sum of the components along the particle's axes. Taking the expectation of this relationship gives the following equation that relates the particle enstrophy, $\left\langle{\boldsymbol{\omega_{p}}}^2\right\rangle$, to the enstrophy about each particle axes:

\begin{equation}
\left\langle{\boldsymbol{\omega_{p}}}^2\right\rangle = \left\langle{\left(\boldsymbol{\omega_{p}} \bcdot \boldsymbol{p_{1}}\right)}^2\right\rangle + \left\langle{\left(\boldsymbol{\omega_{p}} \bcdot \boldsymbol{p_{2}}\right)}^2\right\rangle + 
\left\langle{\left(\boldsymbol{\omega_{p}} \bcdot \boldsymbol{p_{3}}\right)}^2\right\rangle.
\end{equation}

\noindent We refer to the total dimensionless particle enstrophy and its components along the particle's axes as

\begin{figure}
  \centerline{\includegraphics[width=10 cm]{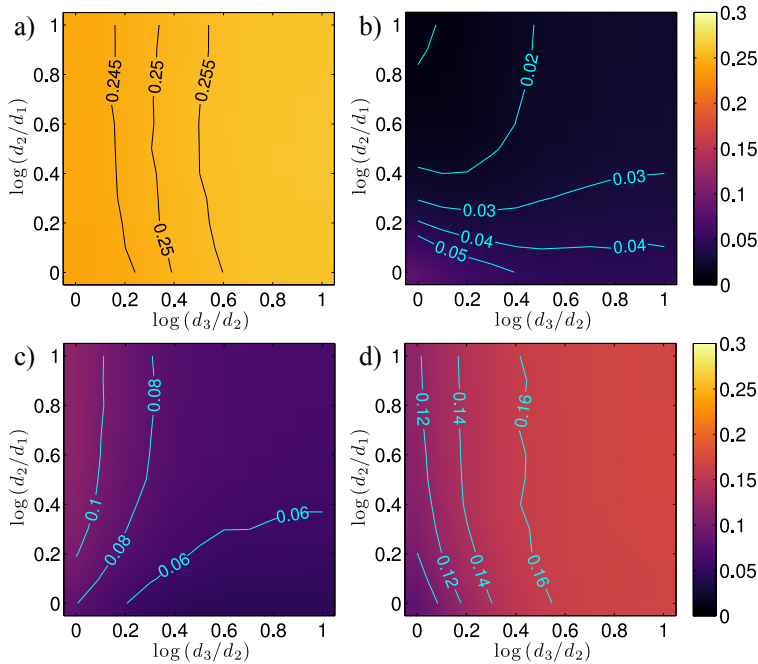}}
  \caption{Variance of particle angular velocities made dimensionless by the Kolmogorov timescale, $\tau_{\eta}$, shown across the particle-shape parameter space: a) $V_{T}$; b) $V_{\boldsymbol{p_{1}}}$; c) $V_{\boldsymbol{p_{2}}}$; d) $V_{\boldsymbol{p_{3}}}$.}
\label{fig:angularvelocity_variance_all}
\end{figure}

\begin{subeqnarray*}
V_{T} = \left\langle{(\boldsymbol{\omega_{p}})}^2\right\rangle \tau_{\eta}^2, \\
V_{\boldsymbol{p_{1}}} = \left\langle{(\boldsymbol{\omega_{p}} \bcdot \boldsymbol{p_{1}})}^2\right\rangle \tau_{\eta}^2, \\ 
V_{\boldsymbol{p_{2}}} = \left\langle{(\boldsymbol{\omega_{p}} \bcdot \boldsymbol{p_{2}})}^2\right\rangle \tau_{\eta}^2, \\ 
V_{\boldsymbol{p_{3}}} = \left\langle{(\boldsymbol{\omega_{p}} \bcdot \boldsymbol{p_{3}})}^2\right\rangle \tau_{\eta}^2,
\label{eq:particle_enstrophy}
\end{subeqnarray*}

\noindent respectively, and plot the results in Figure \ref{fig:angularvelocity_variance_all}. The most striking result is that the total enstrophy of particles is only very weakly shape-dependent (figure \ref{fig:angularvelocity_variance_all}a), but the estrophy for rotation about different axes show strong shape-dependency (figures \ref{fig:angularvelocity_variance_all}b--d). The shape-dependency shown here replicates and extends results shown by \citet{Parsa12}, \citet{ChevillardMeneveau13}, and \citet{Byron15}. We see that particle enstrophy increases slightly as particles become elongated (increasing $d_{3}/d_{2}$), but not as they become flattened (increasing $d_{2}/d_{1}$). This result is explained in section \ref{sec:rotation_sources}, where the contributions to particle enstrophy due to vorticity and strain-rate are examined separately. In appendix \ref{sec:appA} we examine the results for particle angular momentum, analogous to the those shown in figure \ref{fig:angularvelocity_variance_all} for particle enstrophy, calculated using the moment-of-inertia tensor. Neither the particle enstrophy pattern in figure \ref{fig:angularvelocity_variance_all}a, nor the angular momentum patterns in appendix \ref{sec:appA}, show trends matching those of the shape factors $\Psi$ and $\Phi$ (figure \ref{fig:sphericity}). This indicates that those geometric shape factors are insufficient for describing particle rotation in turbulence.

Figures \ref{fig:angularvelocity_variance_all}b, \ref{fig:angularvelocity_variance_all}c, and \ref{fig:angularvelocity_variance_all}d show that $V_{\boldsymbol{p_{3}}} \ge V_{\boldsymbol{p_{2}}} \ge V_{\boldsymbol{p_{1}}}$ for all shapes. Since the total enstrophy is almost shape-independent, it can be concluded that particles of all shapes are most likely to rotate about $\boldsymbol{p_{3}}$, the longest axis, and least likely to rotate about $\boldsymbol{p_{1}}$, the shortest axis. $V_{\boldsymbol{p_{1}}}$ varies most strongly with $d_{2}/d_{1}$, its value decreasing as $d_{2}/d_{1}$ increases and particles become more flattened. $V_{\boldsymbol{p_{3}}}$ varies most strongly with $d_{3}/d_{2}$, its value increasing as $d_{3}/d_{2}$ increases and particles become more elongated. The bulk of the variation in both cases occurs in the range of aspect ratio $(d_{3}/d_{2}$, $d_{2}/d_{1}) = 1 \rightarrow 2$. 

While \citet{ChevillardMeneveau13} show the rotation \textit{of} the particle axes, we focus on particle rotation \textit{about} each axis. For example, \citet{ChevillardMeneveau13} show that ${\scriptstyle \left\langle{(\dot{\boldsymbol{p_{2}}})}^2\right\rangle \tau_{\eta}^2} = V_{\boldsymbol{p_{1}}} + V_{\boldsymbol{p_{3}}}$ is near-constant along the line $d_{3}/d_{2} = d_{2}/d_{1}$, while the results in figure \ref{fig:angularvelocity_variance_all} show that $V_{\boldsymbol{p_{2}}}$ is also near-constant and that there are systematic variations in $V_{\boldsymbol{p_{1}}}$ and $V_{\boldsymbol{p_{3}}}$ which cancel each other along this line. 

The kurtosis of the particle angular velocity and its components are given by

\begin{subeqnarray*}
K_{T} = {{\left\langle{\boldsymbol{\omega_{p}}}^4\right\rangle}} {{\left\langle{\boldsymbol{\omega_{p}}}^2 \right\rangle}^{-2}}, \\
K_{\boldsymbol{p_{1}}} = {\left\langle{\left(\boldsymbol{\omega_{p}} \bcdot \boldsymbol{p_{1}}\right)}^4\right\rangle}{{\left\langle{\left(\boldsymbol{\omega_{p}} \bcdot \boldsymbol{p_{1}}\right)}^2\right\rangle}^{-2}}, \\ 
K_{\boldsymbol{p_{2}}} = {\left\langle{\left(\boldsymbol{\omega_{p}} \bcdot \boldsymbol{p_{2}}\right)}^4\right\rangle}{{\left\langle{\left(\boldsymbol{\omega_{p}} \bcdot \boldsymbol{p_{2}}\right)}^2\right\rangle}^{-2}},\\ 
K_{\boldsymbol{p_{3}}} = {\left\langle{\left(\boldsymbol{\omega_{p}} \bcdot \boldsymbol{p_{3}}\right)}^4\right\rangle}{{\left\langle{\left(\boldsymbol{\omega_{p}} \bcdot \boldsymbol{p_{3}}\right)}^2\right\rangle}^{-2}},
\label{eq:particle_kurtosis}
\end{subeqnarray*}

\begin{figure}
  \centerline{\includegraphics[width=10 cm]{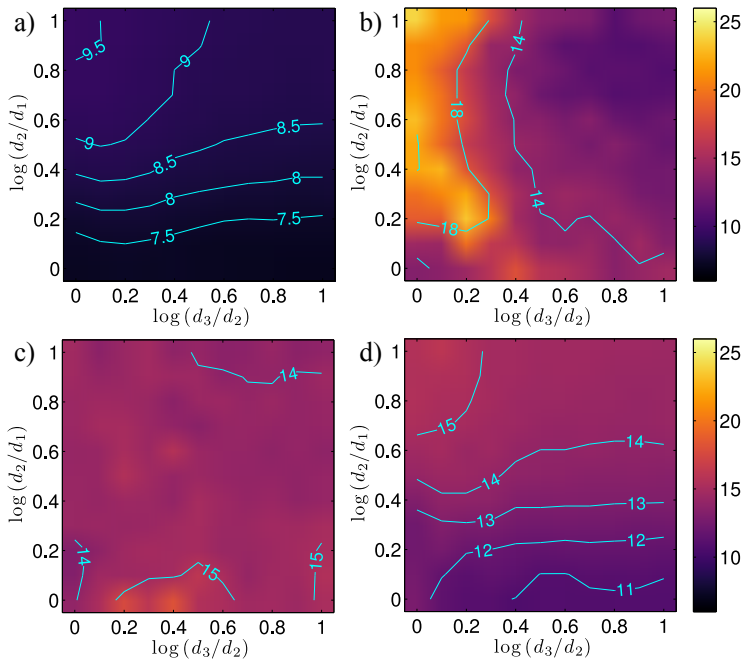}}
  \caption{Kurtosis of particle angular velocities shown across the particle-shape parameter space: a) $K_{T}$; b) $K_{\boldsymbol{p_{1}}}$; c) $K_{\boldsymbol{p_{2}}}$; d) $K_{\boldsymbol{p_{3}}}$.}
\label{fig:angularvelocity_kurtosis_all}
\end{figure}

\noindent and plotted in Figure \ref{fig:angularvelocity_kurtosis_all}. The kurtosis quantifies the likelihood of extreme events. The value of $K_{T}$ falls within 7 and 10, much higher than the value of 3 for a standard normal distribution. $K_{T}$ also varies much more strongly than $V_{T}$ over the shape space, and shows a different dependence on particle shape. That is, while flattening (increasing $d_{2}/d_{1}$) does not affect $V_{T}$, it strongly affects $K_{T}$. In Figure 5, the data contain some noise due to the slow convergence of higher-order moments, but it can be seen that values of $K_{\boldsymbol{p_{i}}}$ for $i=1,2,3$ are larger than those of $K_{T}$ regardless of shape. This is relevant to plankton because their sensory organs (\textit{e.g.}, statocysts) are often aligned so as to sense rotation about a single principal axis \citep{Wells68, Budelmann89}. Sensing in this manner, rather than sensing total angular velocity, gives a more leptokurtic signal, which may be more powerful in guiding behvior. The bulk of the variation in $K_{\boldsymbol{p_{1}}}$ and $K_{\boldsymbol{p_{3}}}$ occurs in the range of aspect ratio $(d_{3}/d_{2}$, $d_{2}/d_{1}) = 1 \rightarrow 2$, which is a common shape range for planktonic organisms. Interestingly, $K_{\boldsymbol{p_{2}}}$ appears to be independent of particle shape.

Consistent with the results of \cite{ChevillardMeneveau13}, the largest value of kurtosis is observed for $K_{\boldsymbol{p_{1}}}$ for particles with $d_{3}/d_{2}=1, d_{2}/d_{1}\gg1$, \textit{i.e.}, discs spinning about their axis of symmetry. Spinning discs have the lowest value of enstrophy (figure \ref{fig:angularvelocity_variance_all}b), which means that kurtosis is largest where variance is smallest. This indicates that rotational motion which is less common on average is still subject to large deviations from its mean value. 

Kurtosis values show a Reynolds number depedency, seen by comparing our data to results in \citet{ChevillardMeneveau13} and \citet{Parsa12}. \citet{ChevillardMeneveau13} observed a value of $K_{\boldsymbol{p_{1}}} \approx 9$ for spinning motion of disc-like particles, whereas we observe a value of $K_{\boldsymbol{p_{1}}} \approx 20$. We also observe a higher value for the kurtosis of the tumbling motion of rod-like particles, \textit{i.e.}, ${\scriptstyle {\left\langle{(\dot{\boldsymbol{p_{3}}})}^4\right\rangle} / {\left\langle{(\dot{\boldsymbol{p_{3}}})}^2\right\rangle}^2} \approx 10.5$ for particles with $d_{3}/d_{2}\gg 1, d_{2}/d_{1}=1$ (data not plotted). \citet{ChevillardMeneveau13} found 6.3 and \citet{Parsa12} found 7.3. Given that the experimental results of \citet{Parsa12} match their DNS results, and that the values of particle enstrophy match between all three studies, it appears that the kurtosis values are a function of Reynolds number. ${\Rey_{\lambda}} \approx 125$ in \citet{ChevillardMeneveau13}, ${\Rey_{\lambda}} \approx 180$ in \citet{Parsa12} and ${\Rey_{\lambda}} \approx 433$ in this study. The kurtosis value seems to increase with Reynolds number, which is consistent with the fact that it is sensitive to the small-scale intermittency in turbulence \citep{Meneveau11}.

\begin{figure}
  \centerline{\includegraphics[width=13 cm]{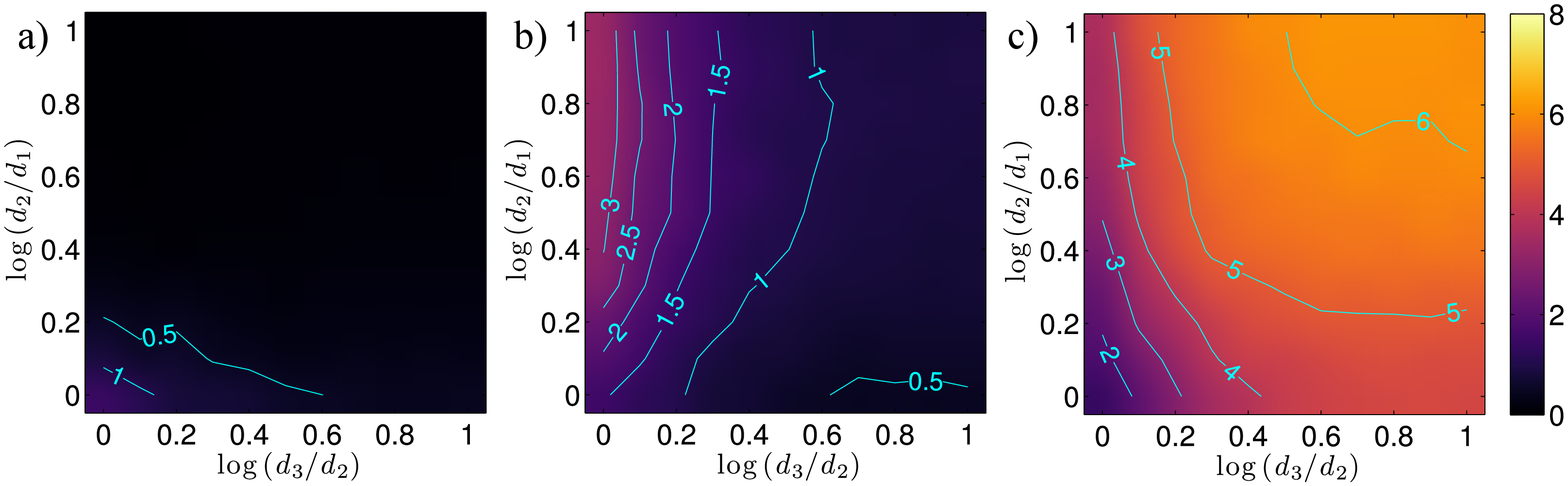}}
  \caption{Fourth moments of particle angular velocities shown across the particle-shape parameter space: a) ${\left\langle{\left(\boldsymbol{\omega_{p}} \bcdot \boldsymbol{p_{1}}\right)}^4\right\rangle}{{\left\langle{\boldsymbol{\omega_{p}}}^2 \right\rangle}^{-2}}$; b) ${\left\langle{\left(\boldsymbol{\omega_{p}} \bcdot \boldsymbol{p_{2}}\right)}^4\right\rangle}{{\left\langle{\boldsymbol{\omega_{p}}}^2 \right\rangle}^{-2}}$; c) ${\left\langle{\left(\boldsymbol{\omega_{p}} \bcdot \boldsymbol{p_{3}}\right)}^4\right\rangle}{{\left\langle{\boldsymbol{\omega_{p}}}^2 \right\rangle}^{-2}}$.}
\label{fig:angularvelocity_fourthmoment_all}
\end{figure}

To isolate shape effects on the fourth moments from shape effects on the second moments, figure \ref{fig:angularvelocity_fourthmoment_all} plots the fourth moments made dimensionless by ${\scriptstyle {\left\langle{\boldsymbol{\omega_{p}}}^2 \right\rangle}^{2}}$. In this case, it can be seen that ${\scriptstyle {\left\langle{\left(\boldsymbol{\omega_{p}} \bcdot \boldsymbol{p_{3}}\right)}^4\right\rangle} \ge {\left\langle{\left(\boldsymbol{\omega_{p}} \bcdot \boldsymbol{p_{2}}\right)}^4\right\rangle} \ge{\left\langle{\left(\boldsymbol{\omega_{p}} \bcdot \boldsymbol{p_{1}}\right)}^4\right\rangle}}$ for all shapes, indicating that extreme angular velocities are most likely to occur about $\boldsymbol{p_{3}}$, the longest axis, and least likely to occur about $\boldsymbol{p_{1}}$, the shortest axis. This is analogous to the trend observed in figure \ref{fig:angularvelocity_variance_all} for the variance of angular velocity. However, in the case of the fourth moment, ${\scriptstyle {\left\langle{\left(\boldsymbol{\omega_{p}} \bcdot \boldsymbol{p_{3}}\right)}^4\right\rangle}}$ increases with both $d_{3}/d_{2}$ and $d_{2}/d_{1}$, and the largest value occurs for very elongated and very flattened triaxial particles ($d_{3}/d_{2} = d_{2}/d_{1} \gg 1$) rather than being common to all elongated particles ($d_{3}/d_{2} \gg 1$). 

We can conclude from the results of this section that the mean magnitude of the particle angular velocity is approximately $0.5{\tau_{\eta}}^{-1}$, and varies weakly with the value of $d_{3}/d_{2}$. Of its three axes, a particle of a given shape is most likely to rotate about its longest axis. These results are independent of Reynolds number, or very nearly so, based on comparison with results of \citet{ShinKoch05}, \citet{Parsa12}, and \citet{ChevillardMeneveau13}. A particle of a given shape is also most likely to experience large angular velocities ($\gg 0.5{\tau_{\eta}}^{-1}$) about its longest axis, and this likelihood increases with values of both $d_{3}/d_{2}$ and $d_{2}/d_{1}$ so that very elongated and very flattened particles are the most likely to experience large angular velocities. The magnitudes of the large angular velocities will depend on Reynolds number, but the trend with particle shape will remain the same. 

\subsection{Particle rotation due to local vorticity and strain-rate \label{sec:rotation_sources}}

\begin{figure}
  \centerline{\includegraphics[width=13 cm]{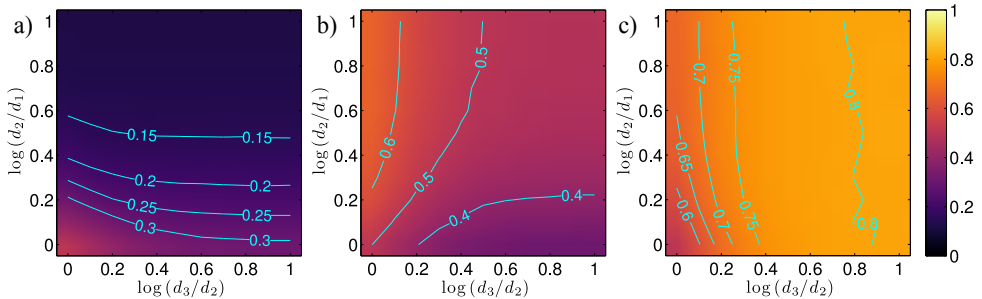}}
  \caption{Alignment of particles' axes with the direction of local fluid vorticity shown across the particle-shape parameter space and quantified by the median value of $|\boldsymbol{e_{\omega}}\bcdot\boldsymbol{p_{i}}|$: a) median of $|\boldsymbol{e_{\omega}}\bcdot\boldsymbol{p_{1}}|$; b) median of $|\boldsymbol{e_{\omega}}\bcdot\boldsymbol{p_{2}}|$; c) median of $|\boldsymbol{e_{\omega}}\bcdot\boldsymbol{p_{3}}|$.}
\label{fig:alignment_vorticity}
\end{figure}

\noindent The trends discussed in the previous section arise due to nonrandom particle orientation with respect to the local velocity gradient tensor. In figure \ref{fig:alignment_vorticity}, the alignment between the particle's axes, $(\boldsymbol{p_{1}}, \boldsymbol{p_{2}}, \boldsymbol{p_{3}})$, and the direction of the local fluid vorticity, $\boldsymbol{e_{\omega}} = \boldsymbol{\omega}/|\boldsymbol{\omega}|$, is shown. Alignment is computed as the median of the distribution of $|\boldsymbol{e_{\omega}}\bcdot\boldsymbol{p_{i}}|$ for $i=1,2,3$.  The probability density functions (PDFs) of $|\boldsymbol{e_{\omega}}\bcdot\boldsymbol{p_{i}}|$ have peaks at 1 or 0; otherwise they are flat. Thus, a median value close to 1 demonstrates strong tendency to be aligned, a median value close to 0 demonstrates strong tendency to be orthogonal, and a median value close to 0.5 demonstrates random alignment. For all shapes, there is strong alignment between $\boldsymbol{p_{3}}$ and vorticity. This tendency becomes even stronger as $d_{3}/d_{2}$ increases and particles become more elongated. For all shapes, there is also a strong tendency for the vorticity and $\boldsymbol{p_{1}}$ to be orthogonal. This tendency becomes stronger as $d_{2}/d_{1}$ increases and particles become more flattened. The alignment between $\boldsymbol{p_{2}}$ and vorticity is almost random for all shapes. 

Particle alignment with vorticity is caused by strain-rate-induced rotations, as can be seen from Eq. (\ref{eq:jeffery}) and Eq. (\ref{eq:omega_p}). These rotations cause the initially randomly orientated particles to approach a statistically steady state of nontrivial alignment. The alignment between $\boldsymbol{p_{3}}$ and vorticity explains the high values of enstrophy for rotations about $\boldsymbol{p_{3}}$ for elongated particles seen in figure \ref{fig:angularvelocity_variance_all}d. The orthogonality between $\boldsymbol{p_{1}}$ and vorticity explains the low values of enstrophy for rotations about $\boldsymbol{p_{1}}$ for flattened particles seen in figure \ref{fig:angularvelocity_variance_all}b. This tendency of elongated particles to be aligned with the fluid vorticity is well-known \citep{ShinKoch05, PumirWilkinson11, Parsa12, ChevillardMeneveau13, Byron15} and is explained by the similarities between the equations governing evolution of vorticity, material lines, and elongated particles \citep{GirimajiPope90, PumirWilkinson11}. The results here show that the same effect occurs for all elongated particles ($d_{3}/d_{2} \gg 1$) regardless of how flattened they are. The degree to which particles are flattened determines the further sub-partitioning of the remaining enstrophy between axes $\boldsymbol{p_{1}}$ and $\boldsymbol{p_{2}}$ (figures \ref{fig:angularvelocity_variance_all}b,c).

Apart from causing particles to align with the vorticity, strain-rate-induced rotations also directly affect particle enstrophy. The separate contributions from vorticity and strain-rate can be examined by taking the expectation of the square of Eq. (\ref{eq:jeffery}) to give the following equations:

\begin{subeqnarray}
  \left\langle\left(\boldsymbol{\omega_{p}}\bcdot\boldsymbol{p_{1}}\right)^2\right\rangle = \frac{1}{4}\left\langle\left(\boldsymbol{\omega}\bcdot\boldsymbol{p_{1}}\right)^2\right\rangle + \left\langle\lambda_{1}^2\left( {\boldsymbol{p_{2}}}^{T}\boldsymbol{S}\boldsymbol{p_{3}}\right)^2\right\rangle + \left\langle\lambda_{1}\left(\boldsymbol{\omega_{p}}\bcdot\boldsymbol{p_{1}}\right)\left( {\boldsymbol{p_{2}}}^{T}\boldsymbol{S}\boldsymbol{p_{3}}\right)\right\rangle, \\
\left\langle\left(\boldsymbol{\omega_{p}}\bcdot\boldsymbol{p_{2}}\right)^2\right\rangle = \frac{1}{4}\left\langle\left(\boldsymbol{\omega}\bcdot\boldsymbol{p_{2}}\right)^2\right\rangle + \left\langle\lambda_{2}^2\left( {\boldsymbol{p_{3}}}^{T}\boldsymbol{S}\boldsymbol{p_{1}}\right)^2\right\rangle + \left\langle\lambda_{2}\left(\boldsymbol{\omega_{p}}\bcdot\boldsymbol{p_{2}}\right)\left( {\boldsymbol{p_{3}}}^{T}\boldsymbol{S}\boldsymbol{p_{1}}\right)\right\rangle, \\
\left\langle\left(\boldsymbol{\omega_{p}}\bcdot\boldsymbol{p_{3}}\right)^2\right\rangle = \frac{1}{4}\left\langle\left(\boldsymbol{\omega}\bcdot\boldsymbol{p_{3}}\right)^2\right\rangle + \left\langle\lambda_{3}^2\left( {\boldsymbol{p_{1}}}^{T}\boldsymbol{S}\boldsymbol{p_{2}}\right)^2\right\rangle + \left\langle\lambda_{3}\left(\boldsymbol{\omega_{p}}\bcdot\boldsymbol{p_{3}}\right)\left( {\boldsymbol{p_{1}}}^{T}\boldsymbol{S}\boldsymbol{p_{2}}\right)\right\rangle.
\label{eq:enstrophy_decomposition}
\end{subeqnarray}

\noindent In the equations above, the contribution to enstrophy solely from fluid vorticity is given by the first term, the contribution to enstrophy solely from strain-rate is given by the second term, and the third term gives the contribution from the cross-correlation of vorticity-induced rotations and strain-rate-induced rotations. 

\begin{figure}
  \centerline{\includegraphics[width=7 cm]{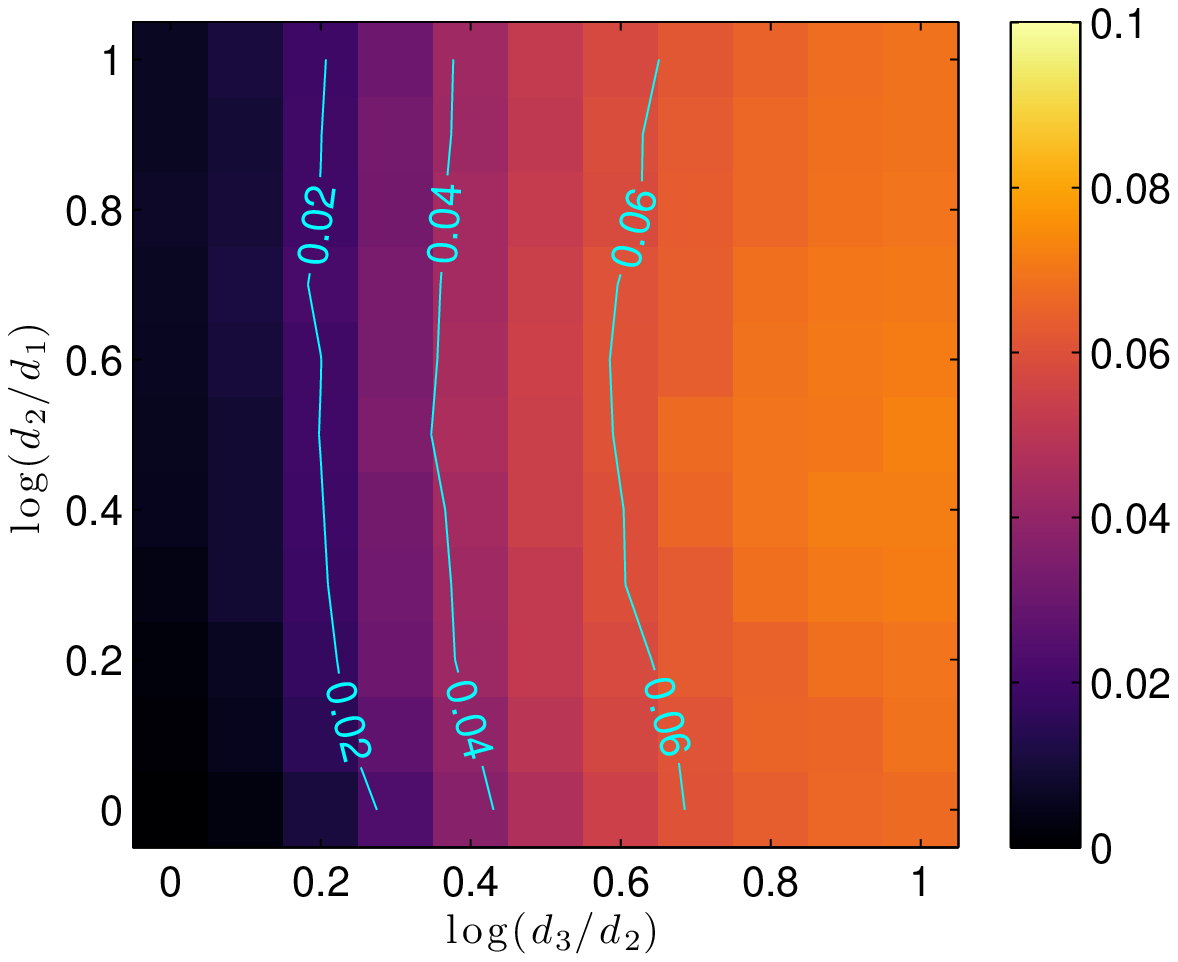}}
\caption{Fraction of particle enstrophy due to strain-rate-induced rotations shown across the particle-shape parameter space: ${\scriptstyle \left(\left\langle{\boldsymbol{\omega_{p}}}^2\right\rangle - \frac{1}{4}\left\langle\boldsymbol{\omega}^2\right\rangle \right)\left(\left\langle{(\boldsymbol{\omega_{p}})}^2\right\rangle\right)^{-1}}$.}
\label{fig:enstrophy_contributions_strain}
\end{figure}

Figure \ref{fig:enstrophy_contributions_strain} plots the contribution of the second and third terms in Eq. (\ref{eq:enstrophy_decomposition}) to the total particle enstrophy. The variation with shape in figure \ref{fig:enstrophy_contributions_strain} exactly matches that in figure \ref{fig:angularvelocity_variance_all}a, which is to be expected since shape effects are only present in the second and third terms in Eq. (\ref{eq:enstrophy_decomposition}. This serves to demonstrate that strain-rate-induced rotations are the cause of the the weak shape-dependency in total particle enstrophy.

\begin{figure}
  \centerline{\includegraphics[width=10 cm]{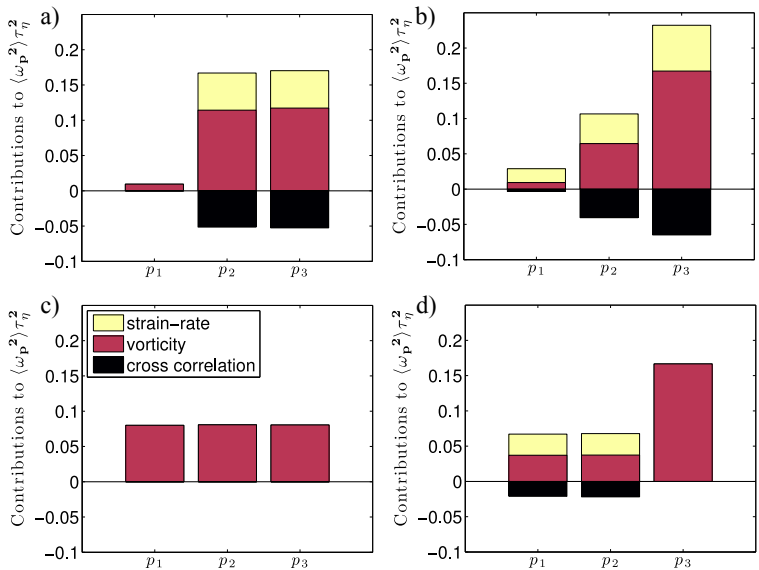}}
  \caption{Dimensionless particle enstrophy about each axis due to vorticity, strain-rate, and cross-correlation between vorticity and strain-rate: a) disc ($d_{3}/d_{2}=1,d_{2}/d_{1}=10$); b) triaxial ($d_{3}/d_{2}=d_{2}/d_{1}=10$); c) sphere ($d_{3}/d_{2}=d_{2}/d_{1}=1$); d) rod ($d_{3}/d_{2}=10,d_{2}/d_{1}=1$). As expected, the results show that axisymmetric particles ($d_{3}/d_{2}=1$ or $d_{2}/d_{1}=1$) can only be spun about their axis of symmetry by fluid vorticity and therefore strain-rate-induced rotations only contribute to rotations \textit{of} the axis of symmetry, \textit{i.e.}, tumbling motions.}
\label{fig:enstrophy_contributions_fourcorners}
\end{figure}

\begin{figure}
  \centerline{\includegraphics[width=13 cm]{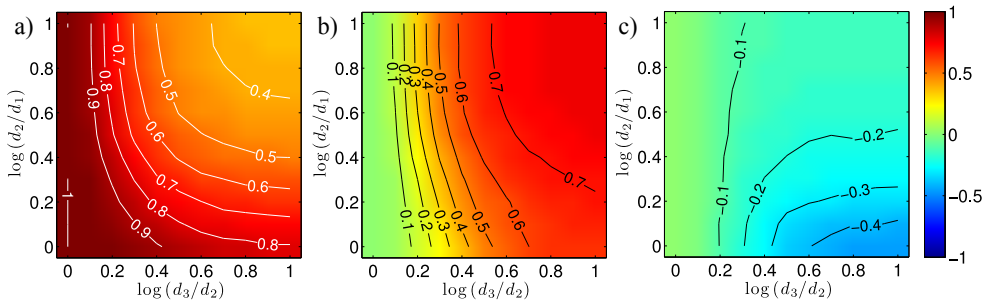}}
\caption{Fraction of particle enstrophy for rotations about $\boldsymbol{p_{1}}$ shown across the particle-shape parameter space: a) due to vorticity alone; b) due to strain-rate alone; c) due to cross-correlation between local vorticity and local strain-rate. The superposition of plots a, b, and c gives a value of 1.}
\label{fig:enstrophy_contributions_p1}
\end{figure}

\begin{figure}
  \centerline{\includegraphics[width=13 cm]{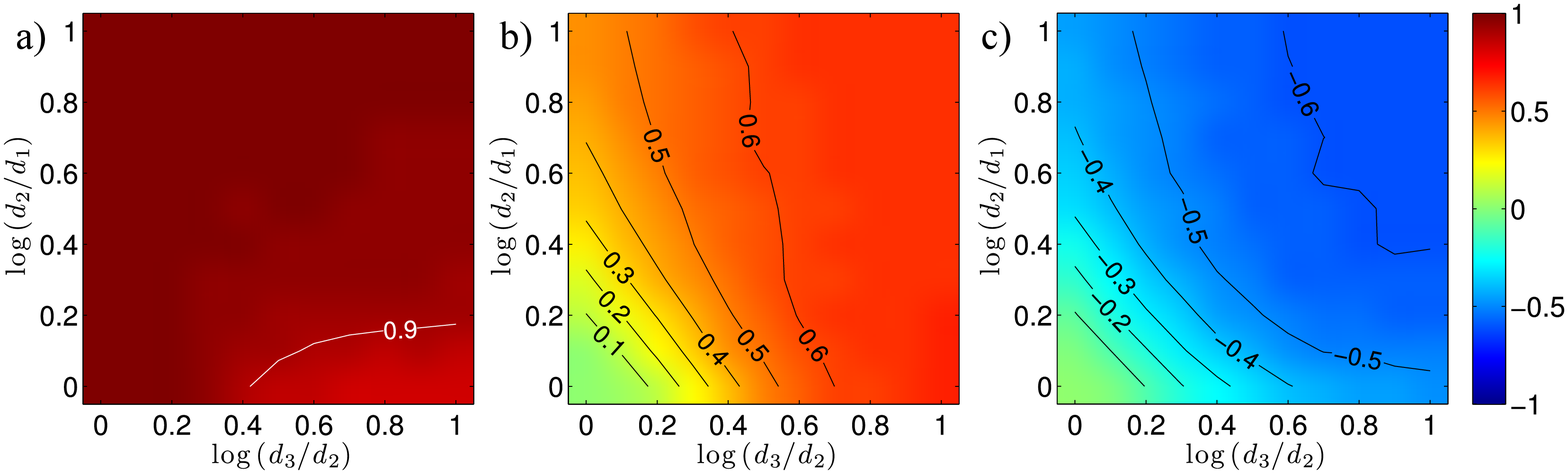}}
\caption{Fraction of particle enstrophy for rotations about $\boldsymbol{p_{2}}$ shown across the particle-shape parameter space: a) due to vorticity alone; b) due to strain-rate alone; c) due to cross-correlation between local vorticity and local strain-rate. The superposition of plots a, b, and c gives a value of 1.}
\label{fig:enstrophy_contributions_p2}
\end{figure}

\begin{figure}
  \centerline{\includegraphics[width=13 cm]{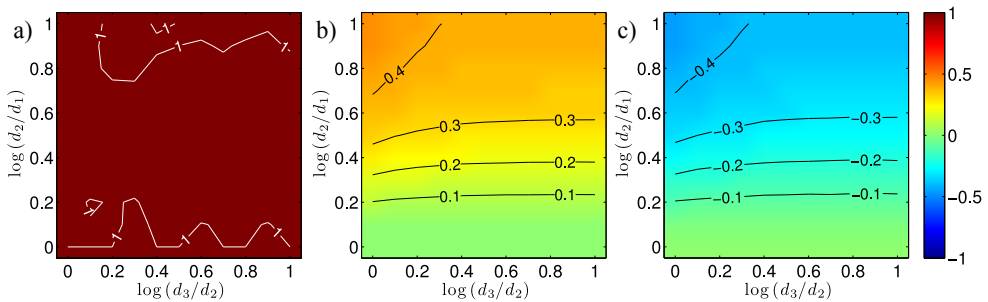}}
\caption{Fraction of particle enstrophy for rotations about $\boldsymbol{p_{3}}$ shown across the particle-shape parameter space: a) due to vorticity alone; b) due to strain-rate alone; c) due to cross-correlation between local vorticity and local strain-rate. The superposition of plots a, b, and c gives a value of 1.}
\label{fig:enstrophy_contributions_p3}
\end{figure}

To explain the pattern seen in figures \ref{fig:angularvelocity_variance_all}a and \ref{fig:enstrophy_contributions_strain}, we examine the effect of each term in Eq. (\ref{eq:enstrophy_decomposition}) on the enstrophy about each particle axis. Figure \ref{fig:enstrophy_contributions_fourcorners} shows the contributions of each term for particles which occupy the four corners of the particle-shape parameter space, whereas figures \ref{fig:enstrophy_contributions_p1}, \ref{fig:enstrophy_contributions_p2}, and \ref{fig:enstrophy_contributions_p3} plot the fractional contributions from each term across the particle-shape paramter space. The data show that vorticity-induced rotations and strain-rate-induced rotations are anticorrelated. This can be seen in figures \ref{fig:enstrophy_contributions_p1}c, \ref{fig:enstrophy_contributions_p2}c, and \ref{fig:enstrophy_contributions_p3}c, and by the darkest bars in figure \ref{fig:enstrophy_contributions_fourcorners}. This shows that along Lagrangian trajectories in turbulence, particle angular velocities due to strain-rate oppose those due to vorticity, on average. The case of strain-rate working against vorticity is familiar from Jeffery's orbits in simple shear flow, where it causes some phases of the Jeffery's orbit to be slower than others.

In examining the relative importance of vorticity-induced rotations and strain-rate-induced rotations to the tumbling motion of rods (${\scriptstyle {\left\langle{(\dot{\boldsymbol{p_{3}}})}^2\right\rangle}\tau_{\eta}^2 = V_{\boldsymbol{p_{1}}} + V_{\boldsymbol{p_{2}}}}$ for $d_{3}/d_{2} \gg 1$ and $d_{2}/d_{1}=1$), \citet{Ni15} observed that the variance of rod tumbling conditioned on local value of fluid enstrophy, ${\boldsymbol{\omega}}^2$, and dissipation, $2\nu S_{ij}S_{ij}$, showed the tumbling rate monotonically increased with both fluid enstrophy and dissipation. Thus, they concluded that vorticity-induced rotations and strain-rate-induced rotations contribute equally to the tumbling of rods. However, examining the bottom right corners of the plots in figures \ref{fig:enstrophy_contributions_p1} and \ref{fig:enstrophy_contributions_p2}, we see that vorticity is responsible for approximately 80\% of ${V_{\boldsymbol{p_{1}}}}$ and ${V_{\boldsymbol{p_{2}}}}$. \citeauthor{Ni15}'s results can be explained by the fact that the fluid enstrophy and dissipation are positively correlated along Lagrangian trajectories \citep[data not plotted, see also][]{Zeff03}.

We noted from figures \ref{fig:angularvelocity_kurtosis_all} and \ref{fig:angularvelocity_fourthmoment_all} that extreme angular velocities are most likely to occur about the longest axis, $\boldsymbol{p_{3}}$, and that larger extremes are more likely as $d_{2}/d_{1}$ increases and particles become flattened. To explore the cause of this trend, we isolate extreme particle angular velocities by only considering those events whose probability of occuring is less than or equal to $0.1\%$. The magnitude of these extreme particle angular velocities is $\langle\left(\boldsymbol{\omega_{p}}|P_{i} \ge 99.9 \right)^{2}\rangle = 48\langle\boldsymbol{\omega_{p}}^{2}\rangle$ averaged across particles of all shapes. The average value of fluid enstrophy and dissipation during these extreme particle angular velocity events is $\langle\left(\boldsymbol{\omega}|P_{i} \ge 99.9\right)^2\rangle = 38\langle\boldsymbol{\omega}^2\rangle$ and $\langle\left(\epsilon|P_{i} \ge 99.9 \right)\rangle = 9\langle\epsilon\rangle$, respectively. This confirms that particles experience large angular velocities in regions of the flow where fluid enstrophy and dissipation are both much higher than their average values. 

\begin{figure}
  \centerline{\includegraphics[width=13 cm]{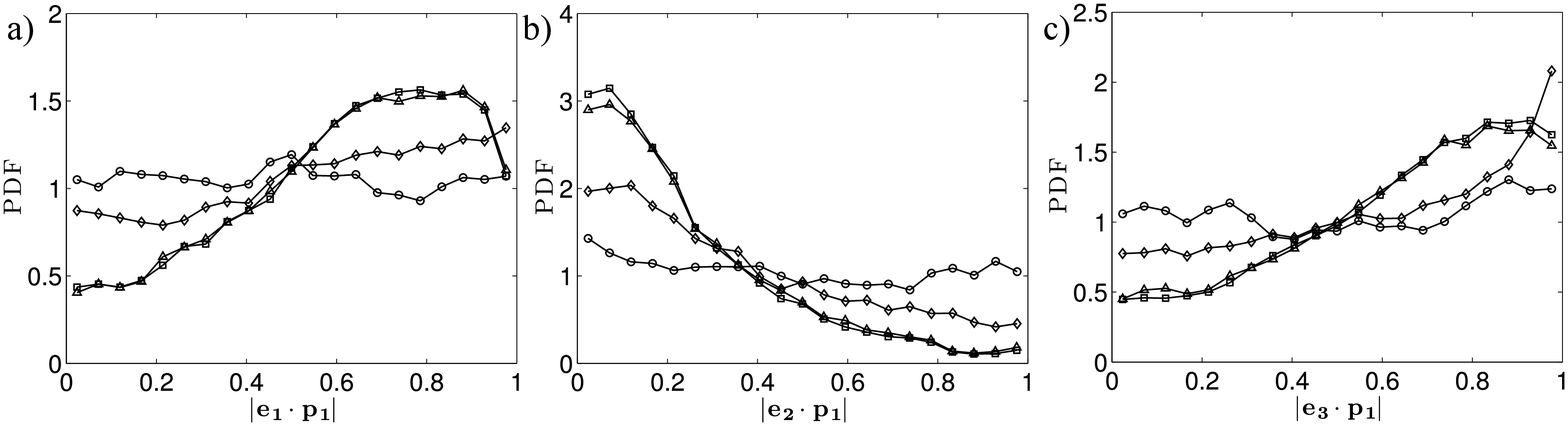}}
  \caption{PDF of alignment between the shortest particle axis and the strain-rate eigen-directions during extreme particle angular velocities: a) $|\boldsymbol{e_{1}}\bcdot\boldsymbol{p_{1}}|$; b) $|\boldsymbol{e_{2}}\bcdot\boldsymbol{p_{1}}|$; c) $|\boldsymbol{e_{3}}\bcdot\boldsymbol{p_{1}}|$. \protect$\circ$, sphere ($d_{3}/d_{2}=d_{2}/d_{1}=1$); \protect$\square$, disc ($d_{3}/d_{2}=1,d_{2}/d_{1}=10$); \protect$\diamond$, rod ($d_{3}/d_{2}=10,d_{2}/d_{1}=1$); \protect$\triangle$, triaxial ($d_{3}/d_{2}=d_{2}/d_{1}=10$).}
\label{fig:extremesPDF_alignment}
\end{figure}

During extreme particle angular velocity events, the pattern of alignment is different than that shown in figure \ref{fig:alignment_vorticity}. Figure \ref{fig:extremesPDF_alignment} shows the PDF of the alignment between the shortest particle axis, $\boldsymbol{p_{1}}$, and the eigen-directions of strain-rate conditioned on extreme particle angular velocities for particles that occupy the four corners of the particle-shape parameter space. The eigen-directions of strain-rate are labelled such that $\boldsymbol{e_{1}}$, $\boldsymbol{e_{2}}$, $\boldsymbol{e_{3}}$ correspond to the most extensional, intermediate, and most compressional directions, respectively. We choose to show the alignment of the shortest particle axis with the eigen-directions of strain-rate because the shortest axis is expected to show the strongest trends as particles become flattened. To explain figure \ref{fig:extremesPDF_alignment}, we first note that since extreme angular velocities are most likely to occur about $\boldsymbol{p_{3}}$ due to alignment with vorticity and since vorticity tends to be aligned with $\boldsymbol{e_{2}}$ \citep{Ashurst87, Meneveau11, Ni14}, we expect that the shortest particle axis will be orthogonal to $\boldsymbol{e_{2}}$ for anisotropic particles. This is indeed observed in figure \ref{fig:extremesPDF_alignment}b. Figures \ref{fig:extremesPDF_alignment}a and \ref{fig:extremesPDF_alignment}c show that the most likely orientation of flattened particles ($d_{2}/d_{1} \gg 1$) during extreme angular velocities is such that the shortest axis bisects the plane spanned by $\boldsymbol{e_{1}}$ and $\boldsymbol{e_{3}}$, forming a 45-degree angle with both. In this orientation, $|\boldsymbol{e_{1}}\bcdot\boldsymbol{p_{1}}|=|\boldsymbol{e_{3}}\bcdot\boldsymbol{p_{1}}|=\cos{(\pi/4)}\approx0.7$. As discussed by \citet{Ni15} in the context of rotations of rods, the strain-rate-induced rotations are maximised in this orientation. Thus, the alignment of flattened particles with the strain-rate tensor further increases the magnitudes of their extreme angular velocities compared to particles that are axisymetric about their longest axis. 

\begin{figure}
  \centerline{\includegraphics[width=13 cm]{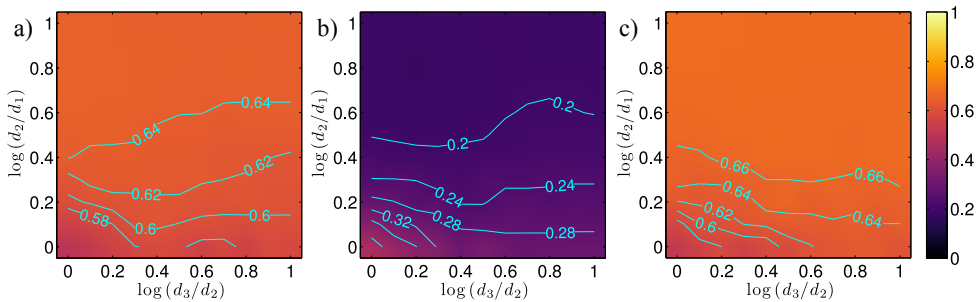}}
  \caption{Alignment of particle's shortest axis with the strain-rate eigen-directions during extreme particle angular velocities shown across the particle-shape parameter space: a) median of $|\boldsymbol{e_{1}}\bcdot\boldsymbol{p_{1}}|$; b) median of $|\boldsymbol{e_{2}}\bcdot\boldsymbol{p_{1}}|$; c) median of $|\boldsymbol{e_{3}}\bcdot\boldsymbol{p_{1}}|$.}
\label{fig:extremes_alignment}
\end{figure}

Figure \ref{fig:extremes_alignment} shows the alignment of the shortest axis with the strain-rate tensor during extreme particle angular velocities across the entire particle-shape parameter space via the median values of $|\boldsymbol{e_{i}}\bcdot\boldsymbol{p_{1}}|$ for $i=1,2,3$. As expected, increasing $d_{2}/d_{1}$ alters the particle orientation during extreme angular velocities and moves it towards the orientation where $\boldsymbol{p_{1}}$ and $\boldsymbol{p_{2}}$ both bisect the plane spanned by the most extensional and the most contracting strain-rate directions. Thus, the source of higher values of the fourth moment of flattened particles ($d_{2}/d_{1} \gg 1$) seen in figure \ref{fig:angularvelocity_kurtosis_all}a is the addition of strain-rate-induced rotations to the already-high values of vorticity-induced rotations about the longest axis in regions of the flow where fluid enstrophy and dissipation are both large. 

Computing the alignment of the shortest axis without conditioning on extreme angular velocities shows a different alignment with the strain-rate tensor; the shortest axis is most commongly aligned with $\boldsymbol{e_{3}}$ \citep[data not plotted here, but consistent with findings of][]{ChevillardMeneveau13}. This also explains why the peak in the PDF of $|\boldsymbol{e_{3}}\bcdot\boldsymbol{p_{1}}|$ in figure \ref{fig:extremesPDF_alignment}c shows a bias towards higher values than the peak in the PDF of $|\boldsymbol{e_{1}}\bcdot\boldsymbol{p_{1}}|$ in figure \ref{fig:extremesPDF_alignment}a.

We can conclude from the results of this section that the majority of particle enstrophy is due to vorticity-induced rotations, but the weak increase in particle enstrophy with the value of $d_{3}/d_{2}$ is due to the addition of strain-rate-induced rotations about $\boldsymbol{p_{1}}$ and $\boldsymbol{p_{2}}$. The largest share of particle enstrophy is in rotations about the longest axis, $\boldsymbol{p_{3}}$, due to its alignment with vorticity. The increase in the kurtosis of particle angular velocity with the value of $d_{2}/d_{1}$ is due to contributions of strain-rate-induced rotations about $\boldsymbol{p_{3}}$. The extremely large particle angular velocities occur when fluid enstrophy and dissipation are both large and the particle orientation maximises strain-rate-induced rotations. 

\subsection{Persistence of particle rotations}

\noindent To understand the timescale of particle rotation and how it relates to timescales of turbulent motions at different scales, we exame the two-time autocorrelation function of particles' angular velocity. This autocorrelation function about axis $\boldsymbol{p_{i}}$ for $i=1,2,3$ is given by

\begin{equation}
\rho_{\left(\boldsymbol{\omega_{p}}\bcdot\boldsymbol{p_{i}}\right)} (s) = \frac{\left\langle{\left(\boldsymbol{\omega_{p}}(t)\bcdot\boldsymbol{p_{i}}(t)\right) \left(\boldsymbol{\omega_{p}}(t+s)\bcdot\boldsymbol{p_{i}}(t+s)\right)}\right\rangle} {\left\langle\left(\boldsymbol{\omega_{p}}\bcdot\boldsymbol{p_{i}}\right)^2\right\rangle}.
\label{eq:autocorrelation_function}
\end{equation}

\noindent The autocorrelation functions do not follow a simple exponential decay. This is consistent with the findings of \citet{ChevillardMeneveau13}, who showed that statistics of particle rotation could not be reproduced with a stochastic model for the velocity gradient tensor that followed a linear Ornstein-Uhlenbeck process, and \citet{ShinKoch05}, who showed that autocorrelation functions do not follow an exponential decay for fibre rotations. 

\begin{figure}
  \centerline{\includegraphics[width=13 cm]{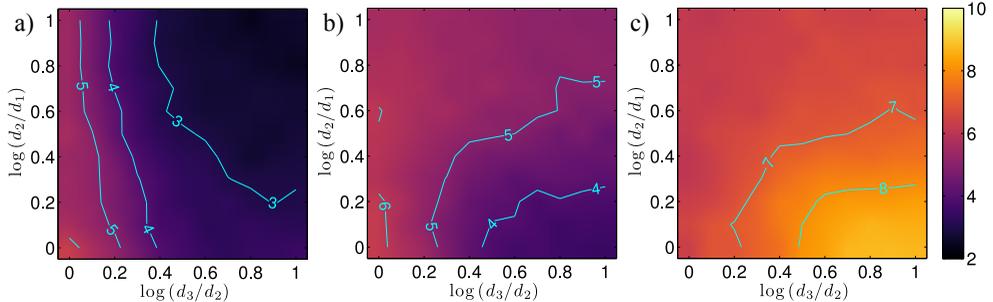}}
  \caption{Dimensionless integral timescale for particle angular velocity shown across the particle-shape parameter space: a) $T_{\boldsymbol{p_{1}}}{\tau_{\eta}}^{-1}$; b) $T_{\boldsymbol{p_{2}}}{\tau_{\eta}}^{-1}$; c) $T_{\boldsymbol{p_{3}}}{\tau_{\eta}}^{-1}$.}
\label{fig:integral_timescale}
\end{figure}

The autocorrelation functions can be used to quantify the timescale over which particle angular velocity is persistent about different axes. This is done by computing an integral timescale for particle angular velocity, which is given by the integral of the autocorrelation function \citep{Pope00}:

\begin{equation}
T_{\boldsymbol{p_{i}}} = \int_{0}^{\infty} \rho_{\left(\boldsymbol{\omega_{p}}\bcdot\boldsymbol{p_{i}}\right)} (s) ds,
\label{eq:integral_timescale}
\end{equation}

\noindent for particle angular velocity about axis $\boldsymbol{p_{i}}$ for $i=1,2,3$. The integral timescales are shown across the particle-shape parameter space in figure \ref{fig:integral_timescale}. The trends show that for all particle shapes, rotations about $\boldsymbol{p_{3}}$ are not only the most likely (figure \ref{fig:angularvelocity_variance_all}), but also the most persistent, lasting between 5 and 10 multiples of $\tau_{\eta}$. Conversely, rotations about $\boldsymbol{p_{1}}$ are the least likely (figure \ref{fig:angularvelocity_variance_all}) and are also the least persistent, lasting between 2 and 5 multiples of $\tau_{\eta}$. Figure \ref{fig:enstrophy_contributions_p3} shows that only vorticity-induced rotations are responsible for the particle angular velocity about $\boldsymbol{p_{3}}$. Therefore, the longer persistence of rotations about $\boldsymbol{p_{3}}$ is consistent with the persistence of vorticity at the smallest scales \citep{SreenivasanAntonia97}. Figure \ref{fig:enstrophy_contributions_p1} showed that strain-rate-induced rotations make significant contribution to rotations about $\boldsymbol{p_{1}}$, especially for shapes that are towards the top right of the particle-shape parameter space. In this case, the shorter persistence of rotations about $\boldsymbol{p_{1}}$ (figure \ref{fig:integral_timescale}a) is consistent with the fact that strain-rate is not as persistent as vorticity at the smallest scales \citep{GirimajiPope90, Pope90}.

Figure \ref{fig:integral_timescale}a shows a similar trend to figure \ref{fig:enstrophy_contributions_p1}a, suggesting that the shape-variation of the integral timescale for rotations about the shortest axis is linked to the relatively contribution of vorticity-induced rotations about this axis. Similarly, figure \ref{fig:integral_timescale}b shows a similar trend to figure \ref{fig:enstrophy_contributions_p2}a, suggesting that the variation in integral timescale for rotations about the intermediate axis is also linked to the relative contribution of vorticity-induced rotations about the intermediate axis. The shape-variation in the integral timescale for rotations about the longest axis in figure \ref{fig:integral_timescale}c are due to a more subtle effect of strain-rate-induced rotations. We see that as elongated particles become flattened ($d_{3}/d_{2} \gg 1$, increasing $d_{2}/d_{1}$), the integral timescale increases. At the same time, from figure \ref{fig:enstrophy_contributions_p3}a, we see that practically all rotation about the longest axis is due to vorticity and from figure \ref{fig:alignment_vorticity}c, we see that there is no difference in the median alignment between $\boldsymbol{p_{3}}$ and vorticity for all elongated particles regardless of how flattened they are ($d_{3}/d_{2} \gg 1$). This indicates that although the overall alignment of the longest axis with vorticity is the same for all elongated particles, it is more persistent for elongated particles that are axisymmetric. In other words, elongated particles that are axisymmetric ($d_{3}/d_{2} \gg1$ and $d_{2}/d_{1} = 1$) maintain alignment with vorticity over a longer timescale than elongated particles that are also flattened ($d_{3}/d_{2} \gg1$ and $d_{2}/d_{1} \gg 1$).

The role played by strain-rate-induced rotations is to maintain alignment of the particle's longest axis with vorticity. But while both the longest particle axis and the vorticity vector tend towards alignment with the most extensional direction of strain-rate when viewed in a Lagrangian sense, particle rotations lead rather than lag the tilting of the vorticity vector \citep{Ni14}. Further, the response of vorticity to strain-rate-induced tilting and stretching is dampened by viscous effects, whereas anisotropic and inertialess particles respond to strain-rate instantaneously and without damping. Additionally, we see from Eq. \ref{eq:jeffery}b and the definition of $\lambda_{2}$ that elongated particles that are also flattened are more sensitive to strain-rate-induced rotations about the middle axis, $\boldsymbol{p_{2}}$, than axisymmetric elongated particles for the same value of $d_{3}/d_{2}$. It is this extra sensitivity to strain-rate that disrupts the alignment of their longest axis with vorticity and leads to lower integral timescales compared to elongated axisymmetric particles.

For plankton, full revolutions about principal axes can be important because they allow for sensing directional environmental cues such as light and gravity \citep{Wells68}, as well as seascape locations \citep{FuchsGerbi16}. Herein, however, we see that such motions are rare in turbulence. The number of revolutions that a particle typically completes about each of its axes before the angular velocity decorrelates with itself can be calculated from the product of the integral timescale and the mean angular velocity. The typical number of revolutions is given by

\begin{equation}
n_{i} = \frac{T_{\boldsymbol{p_{i}}} \sqrt{\left\langle{\left(\boldsymbol{\omega_{p}} \bcdot \boldsymbol{p_{i}}\right)}^2\right\rangle}}{2\pi}.
\label{eq:integral_timescale_revs}
\end{equation}

\begin{figure}
  \centerline{\includegraphics[width=13 cm]{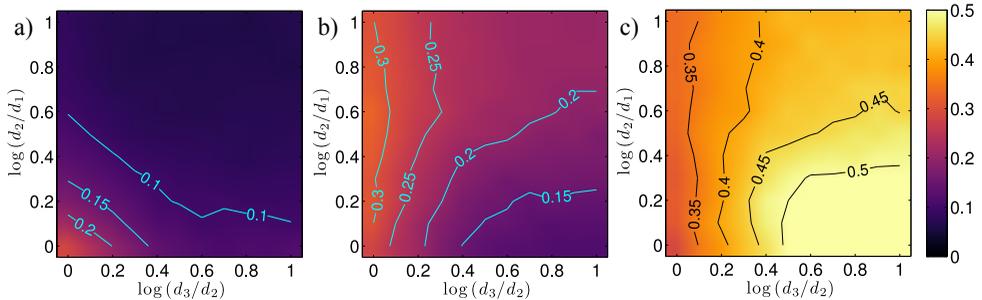}}
  \caption{Number of revolutions completed during an integral timescale for angular velocity across the particle-shape parameter space: a) about $\boldsymbol{p_{1}}$; b) about $\boldsymbol{p_{2}}$; c) about $\boldsymbol{p_{3}}$.}
\label{fig:integral_timescale_revs}
\end{figure}

\noindent The results for $n_{i}$ are plotted in figure \ref{fig:integral_timescale_revs} across the particle-shape parameter space. They show that, typically, less than one-half of a revolution is completed before the angular velocity becomes decorrelated with itself for any particle axis and for any shape. The variation of $n_{i}$ is slightly different than the variation seen in the integral timescale (figure \ref{fig:integral_timescale}) because it also incorporates the variation in the mean angular velocity (figure \ref{fig:angularvelocity_variance_all}).

We can conclude from the results of this section that particle rotations persist for 2--10 multiples of $\tau_{\eta}$ with rotations about the longest axis persisting the longest. This persistence is due to alignment with vorticity being maintained over a longer time than for other axes. Full revolutions about any of the particle axes are rare events, but are most common about the longest axis for axisymmetric elongated particles. The integral timescales of angular velocity depend on intermittency and hence are likely to be a function of Reynolds number, but the trends with particle shape are most likely independent of Reynolds number.

\section{Conclusions}

We have thoroughly examined the rotational dynamics of small, inertialess triaxial ellipsoids in homogenous isotropic turbulence with a focus on describing variations with particle shape after particles have reached statistically-steady alignment with the local velocity gradient tensor. Preferential alignment of particles with the local velocity gradient tensor determines the partitioning of the total particle enstrophy such that the majority of the enstrophy is in rotations about the longest axis. We find that the total particle enstrophy to be a very weak function of shape. This is because vorticity-strain correlations cancel almost all of the strain-rate contributions, so that the bulk of the particle enstrophy is due to fluid vorticity, which acts equally on particles of all shapes. There is a weak increase in total particle enstrophy for elongated particles, which is due to the residual contribution of strain-rate-induced rotations about the particle's shortest and intermediate axes. This means that the slightly higher enstrophy of rods compared to discs observed by \cite{Byron15} is due to additional tumbling motion of rods. 

Particles of all shapes experience angular velocities that are extremely large compared to the mean. These extremely large angular velocities are most likely to occur about the longest axis, and flattened particles are the most likely to experience large extreme angular velocities. These extremes occur when the fluid enstrophy and dissipation are both large.

The integral timescale for particle angular velocities about each axis shows that rotations about the longest axis are the most persistent (linked to the persistence of vorticity at small scales) and rotations about the shortest axis are the least persistent. Events where particles undergo a full revolution about any of their axes are rare, but occur most frequently for axisymmetric rod-like particles for rotations about their axis of symmetry.

Finally, single-parameter descriptions of particle shape, such as the Waddell sphericity parameter and the Corey shape factor, do not to correlate with any of the statistical quantities of particle rotation examined herein. For certain quantities, single-parameter shape descriptions using $d_{3}/d_{2}$ (how elongated a particle is) or $d_{2}/d_{1}$ (how flattened a particle is) are sufficient to describe overall trends. For example, particle enstrophy depends only on $d_{3}/d_{2}$, whereas kurtosis of particle angular velocity depends mainly on $d_{2}/d_{1}$.\\

\noindent Acknowledgements:

The authors would like to acknowledge support from the Army Research Office (Grant no. W911NF-16-1-0284). This research also benefited from the use of the Savio computational cluster resource provided by the Berkeley Research Computing program supported by the University of California, Berkeley. The authors would like to thank Charles Meneveau and Laurent Chevillard for encouragement at the beginning of this project, and Greg Voth for illuminating discussions throughout. The colourmap used in the majority of the figures was obtained from https://bids.github.io/colormap/.

\appendix
\section{Angular momentum and rotational energy}\label{sec:appA}

\noindent In the Stokesian regime in which the equations of particle motion have been derived, particle inertia and moments of inertia are neglected \citep{Jeffery22}. Thus, there is no resultant force or torque on the particle and the particle is at equilibrium with the surrounding fluid at all times. We can nonetheless calculate the particle angular momentum and the particle rotational energy \textit{a posteriori} and examine whether there are any meaningful trends in these quantities. 

The moment of inertia tensor, $\boldsymbol{I}$, for a triaxial ellipsoid is a diagonal matrix in the co-ordinate system fixed to the particle with values along the diagonal, $I_{11},I_{22},I_{33}$, that are the moments of inertia for rotation about the axes $\boldsymbol{p_{1}}, \boldsymbol{p_{2}}, \boldsymbol{p_{3}}$, respectively. The dimensionless moment of inertia tensor is given by

\begin{equation}
\frac{\boldsymbol{I}}{\frac{1}{5}\rho V^{5/3} {\left(\frac{3}{4\pi}\right)}^{2/3}} = \left(\frac{d_{2}/d_{1}}{d_{3}/d_{2}}\right)^{2/3}\left[
\begin{array}{ccc}
  1 + \left(\frac{d_{3}}{d_{2}}\right)^{2}  & 0  &  0  \\
  0  &  \left(\frac{d_{2}}{d_{1}}\right)^{-2} + \left(\frac{d_{3}}{d_{2}}\right)^{2}  &  0 \\
  0  &  0  &  1 + \left(\frac{d_{2}}{d_{1}}\right)^{-2} \\
\end{array}  \right],
\label{eq:dimensionless_moi} 
\end{equation}

\noindent where the values along the diagonal on the right-hand side of Eq. (\ref{eq:dimensionless_moi}) are the dimensionless values of $I_{11},I_{22},I_{33}$, respectively. For particles of all shapes, $I_{11} \ge I_{22} \ge I_{33}$.

Due to particle anisotropy, the total particle angular momentum and rotational energy is not proportional to the varaince of the angular velocity; different moments of inertia about the different axes need to be considered. The total particle angular momentum is the sum of its components along the particle's axes. Taking the expectation of this relationship gives the following equation

\begin{equation}
\left\langle{\left(\boldsymbol{I}\boldsymbol{\omega_{p}}\right)}^2\right\rangle = {I_{11}}^{2}\left\langle{\left(\boldsymbol{\omega_{p}} \bcdot \boldsymbol{p_{1}}\right)}^2\right\rangle + {I_{22}}^{2}\left\langle{\left(\boldsymbol{\omega_{p}} \bcdot \boldsymbol{p_{2}}\right)}^2\right\rangle + 
{I_{33}}^{2}\left\langle{\left(\boldsymbol{\omega_{p}} \bcdot \boldsymbol{p_{3}}\right)}^2\right\rangle,
\label{eq:angular_momentum}
\end{equation}

\noindent which relates the variances of angular velocities about the particle's axes to the variance of the total angular momentum of the particle. A similar equation for the rotational energy gives

\begin{equation}
\left\langle{\frac{1}{2}\boldsymbol{I}\boldsymbol{\omega_{p}}^2}\right\rangle = \frac{1}{2}{I_{11}}\left\langle{\left(\boldsymbol{\omega_{p}} \bcdot \boldsymbol{p_{1}}\right)}^2\right\rangle + \frac{1}{2}{I_{22}}\left\langle{\left(\boldsymbol{\omega_{p}} \bcdot \boldsymbol{p_{2}}\right)}^2\right\rangle + 
\frac{1}{2}{I_{33}}\left\langle{\left(\boldsymbol{\omega_{p}} \bcdot \boldsymbol{p_{3}}\right)}^2\right\rangle.
\label{eq:rotational_energy}
\end{equation}

Since the total particle enstrophy is a weak function of shape, but the moments of inertia about the particle's axes are a strong function of shape, we observe that variations in the variance of particle angular momentum and rotational energy are be dominated by the moments of inertia (data not plotted).

\bibliography{ellipsoid_dynamics}
\bibliographystyle{jfm}

\end{document}